\documentclass[prd,a4paper,twocolumn]{revtex4}
\usepackage{graphicx}
\usepackage{subfigure,amssymb}

\newcommand{\nc}{\newcommand}

\nc{\be}[1]{\begin{equation}\mbox{$\label{#1}$}}
\nc{\bea}[1]{\begin{eqnarray} \mbox{$\label{#1}$}}
\nc{\Section}[2]{\section{#2}\label{#1}}
\nc{\Bibitem}[1]{\bibitem{#1}}
\nc{\Label}[1]{\label{#1}}

\nc{\eea}{\end{eqnarray}}
\nc{\ee}{\end{equation}}

\nc{\bdm}{\begin{displaymath}}
\nc{\edm}{\end{displaymath}}
\nc{\dpsty}{\displaystyle}
\nc{\bc}{\begin{center}}
\nc{\ec}{\end{center}}
\nc{\ba}{\begin{array}}
\nc{\ea}{\end{array}}
\nc{\bab}{\begin{abstract}}
\nc{\eab}{\end{abstract}}
\nc{\btab}{\begin{tabular}}
\nc{\etab}{\end{tabular}}
\nc{\bit}{\begin{itemize}}
\nc{\eit}{\end{itemize}}
\nc{\ben}{\begin{enumerate}}
\nc{\een}{\end{enumerate}}
\nc{\bfig}{\begin{figure}}
\nc{\efig}{\end{figure}}

\nc{\arreq}{&\!=\!&}
\nc{\arrmi}{&\!-\!&}
\nc{\arrpl}{&\!+\!&}
\nc{\arrap}{&\!\!\!\approx\!\!\!&}
\nc{\non}{\nonumber}
\nc{\align}{\!\!\!\!\!\!\!\!&&}

\def\lsim{\; \raise0.3ex\hbox{$<$\kern-0.75em
      \raise-1.1ex\hbox{$\sim$}}\; }
\def\gsim{\; \raise0.3ex\hbox{$>$\kern-0.75em
      \raise-1.1ex\hbox{$\sim$}}\; }

\nc{\DOT}{\hspace{-0.08in}{\bf .}\hspace{0.1in}}
\nc{\Laada}{\hbox {$\sqcap$ \kern -1em $\sqcup$}}
\nc\loota{{\scriptstyle\sqcap\kern-0.55em\hbox{$\scriptstyle\sqcup$}}}
\nc\Loota{{\sqcap\kern-0.65em\hbox{$\sqcup$}}}
\nc\laada{\Loota}
\nc{\qed}{\hskip 3em \hbox{\BOX} \vskip 2ex}

\nc{\real}{{\rm I \! R}}
\nc{\Z}{{\sf Z \!\!\! Z}}
\nc{\complex}{{\rm C\!\!\! {\sf I}\,\,}}
\def\bigid{\leavevmode\hbox{\small1\kern-3.8pt\normalsize1}}
\def\id{\leavevmode\hbox{\small1\kern-3.3pt\normalsize1}}
\nc{\slask}{\!\!\!/}
\nc{\bis}{{\prime\prime}}
\nc{\pa}{\partial}
\nc{\na}{\nabla}
\nc{\ra}{\rangle}
\nc{\la}{\langle}
\nc{\goto}{\rightarrow}
\nc{\swap}{\leftrightarrow}

\nc{\EE}[1]{ \mbox{$\cdot10^{#1}$} }
\nc{\abs}[1]{\left|#1\right|}
\nc{\at}[2]{\left.#1\right|_{#2}}
\nc{\norm}[1]{\|#1\|}
\nc{\abscut}[2]{\Abs{#1}_{\scriptscriptstyle#2}}
\nc{\vek}[1]{{\rm\bf #1}}
\nc{\integral}[2]{\int\limits_{#1}^{#2}}
\nc{\inv}[1]{\frac{1}{#1}}
\nc{\dd}[2]{{{\partial #1}\over{\partial #2}}}
\nc{\ddd}[2]{{{{\partial}^2 #1}\over{\partial {#2}^2}}}
\nc{\dddd}[3]{{{{\partial}^2 #1}\over
    {\partial #2 \partial #3}}}
\nc{\dder}[2]{{{d #1}\over{d #2}}}
\nc{\ddder}[2]{{{d^2 #1}\over{d {#2}^2}}}
\nc{\dddder}[3]{{d^2 #1}\over
    {d #2 d #3}}
\nc{\dx}[1]{d\,^{#1}x}
\nc{\dy}[1]{d\,^{#1}y}
\nc{\dz}[1]{d\,^{#1}z}
\nc{\dl}[1]{\frac{d\,^{#1}l}{(2\pi)^{#1}}}
\nc{\dk}[1]{\frac{d\,^{#1}k}{(2\pi)^{#1}}}
\nc{\dq}[1]{\frac{d\,^{#1}q}{(2\pi)^{#1}}}

\nc{\bfT}{{\bf T }}

\nc{\cA}{{\cal A}}
\nc{\cB}{{\cal B}}
\nc{\cD}{{\cal D}}
\nc{\cE}{{\cal E}}
\nc{\cG}{{\cal G}}
\nc{\cH}{{\cal H}}
\nc{\cL}{{\cal L}}
\nc{\cO}{{\cal O}}
\nc{\cT}{{\cal T}}
\nc{\cN}{{\cal N}}
\nc{\cR}{{\cal R}}
\nc{\rvac}[1]{|{\cal O}#1\rangle}
\nc{\lvac}[1]{\langle{\cal O}#1|}
\nc{\rvacb}[1]{|{\cal O}_\beta #1\rangle}
\nc{\lvacb}[1]{\langle{\cal O}_\beta #1 |}
\nc{\bb}{\bar{\beta}}
\nc{\bt}{\tilde{\beta}}
\nc{\ctH}{\tilde{\cal H}}
\nc{\chH}{\hat{\cal H}}
\nc{\1}{\aa}
\nc{\2}{\"{a}}
\nc{\3}{\"{o}}
\nc{\4}{\AA}
\nc{\5}{\"{A}}
\nc{\6}{\"{O}}
\nc{\al}{\alpha}
\nc{\g}{\gamma}
\nc{\Del}{\Delta}
\nc{\e}{\textrm{e}}
\nc{\eps}{\epsilon}
\nc{\lam}{\lambda}
\nc{\Om}{\Omega}
\nc{\ve}{\varepsilon}
\nc{\mn}{{\mu\nu}}
\nc{\vp}{\varphi}

\nc{\rf}[1]{(\ref{#1})}
\nc{\nn}{\nonumber \\*}
\nc{\bfB}{\bf{B}}
\nc{\bfv}{\bf{v}}
\nc{\bfx}{\bf{x}}
\nc{\bfy}{\bf{y}}
\nc{\vx}{\vec{x}}
\nc{\vy}{\vec{y}}
\nc{\oB}{\overline{B}}
\nc{\oI}{\overline{I}}
\nc{\oR}{\overline{R}}
\nc{\rar}{\rightarrow}
\nc{\ti}{\times}
\nc{\slsh}{\hskip-5pt/}
\nc{\sm}{Standard~Model~}
\nc{\MP}{M_{\rm Pl}}
\nc{\mpl}{M_{\rm Pl}}
\nc{\tp}{t_{\rm Pl}}

\nc{\pmin}{p_{\rm min}}
\nc{\pmax}{p_{\rm max}}
\nc{\fo}{f_0}
\nc{\foi}{f_{0,i}\,}
\nc{\fop}{f_0^P}
\nc{\fou}{f_0^U}

\nc{\eff}{{\rm eff}}
\nc{\MT}{M_{\rm T}}
\nc{\ML}{M_{\rm L}}
\nc{\kk}{\vek{k}}
\nc{\pp}{{\rm p}}
\nc{\pt}{\partial_t}
\nc{\half}{{1\over 2}}
\nc{\w}{\omega}
\nc{\uhat}{\hat{U}_\w}

\nc{\etal}{\mbox{\it et al.}}
\nc{\ie}{{\it i.e. }}
\nc{\eg}{{\it e.g. }}
\nc{\trh}{T_{\rm RH}}
\nc{\ad}{{a'\over a}}
\nc{\bd}{{b'\over b}}
\nc{\Rd}{{R'\over R}}
\nc{\diag}{{\textrm{diag}}}
\nc{\mato}[1]{\tilde{#1}}
\nc{\sech}{\textrm{sech}}
\nc{\I}{\textrm{I}}
\nc{\II}{\textrm{II}}
\nc{\III}{\textrm{III}}
\nc{\vev}[1]{\langle #1 \rangle}
\nc{\hyp}{\,\; F_{1{\hskip -16pt}2}{\hskip 11pt}}
\nc{\brhom}{\overline{\rho}_M}
\nc{\brho}{\overline{\rho}}
\nc{\rhob}{\overline{\rho}}
\nc{\Pb}{\overline{P}}
\nc{\bH}{\overline{H}}
\nc{\ep}{{1+4\eps}}

\nc{\lcdm}{$\Lambda$CDM}

\def\smiley{\hbox{\large$\bigcirc$\hspace{-.80em}%
\raise.2ex\hbox{$\cdot\cdot$}\kern-.61em    
\lower.2ex\hbox{\scriptsize$\smile$}}\ }

\def\frowney{\hbox{\large$\bigcirc$\hspace{-.80em}%
\raise.2ex\hbox{$\cdot\cdot$}\kern-.635em
\lower.2ex\hbox{\scriptsize$\frown$}}\ }

\begin{document}

\title{Constraining Newtonian stellar configurations in $f(R)$ theories of gravity}

\author{T. Multam\"aki}
\email{tuomul@utu.fi}
\author{I. Vilja}
\email{vilja@utu.fi}
\affiliation{Department of Physics, University of Turku, FIN-20014 Turku, FINLAND}

\date{}

\begin{abstract}
We consider general metric $f(R)$ theories of gravity by solving the field equations in the presence of a 
spherical static mass distribution by analytical perturbative means. Expanding the field equations systematically in $\cO(G)$,
we solve the resulting set of equations and show that $f(R)$ theories which attempt to solve the dark energy problem
very generally lead to $\gamma_{PPN}=1/2$ in the solar system. This excludes a large class of theories as 
possible explanations of dark energy. We also present
the first order correction to $\gamma_{PPN}$ and show that it cannot have a significant effect.
\end{abstract}

\maketitle

\section{Introduction}

The dark energy problem remains central in modern day cosmology. Since the matter only, 
homogeneous universe within the framework of general relativity is in conflict with 
cosmological observations, the assumptions behind this model have been questioned. The 
most popular modification is to consider a universe filled with other, more exotic forms of 
matter, the cosmological constant being the leading natural candidate. Other ways to 
tackle the dark energy problem are then to relax the assumption of homogeneity or modify the 
theory of gravity.

In recent years, a particular modification of gravity, the $f(R)$ gravity models that 
replace the Einstein-Hilbert action of general relativity (GR) with an arbitrary function of 
the curvature scalar
(see \eg \cite{turner,turner2,allemandi,meng,nojiri3,nojiri2,cappo1,woodard,odintsov} 
and references therein) have been extensively studied. Naive modification of the 
gravitational action is not without challenges, however, and obstacles including 
cosmological constraints (see \eg \cite{new2,Nojiri:2007as,Starobinsky:2007hu} and references 
therein), 
instabilities \cite{dolgov,soussa,faraoni}, solar system constraints (see {\it e.g.} 
\cite{chiba,confprobs,Clifton,Hu2007} and references therein) and evolution large scale 
perturbations \cite{Bean:2006up,Song:2006ej,Song2} need to be overcome. In addition, a number of 
consistency requirements need to be satisfied (see \eg \cite{Sawicki2007,Appleby} and 
references therein).

One of the most direct and strictest constraints on any modification of gravity comes from 
observations of out nearby space-time \ie the solar system. This is often done by 
conformally transforming the theory to a scalar-tensor theory and then considering the 
Parameterized Post-Newtonian (PPN) limit \cite{damour,magnano} (see also \cite{olmo, ppnok}
for a discussion). The question of validity of the solar system constraints $f(R)$ theories
has been extensively discussed in the literature and not completely without controversy.
The opinions on the viability of $f(R)$ theories have been divided from more or less skeptical 
\cite{Erickcek:2006vf,Chiba2,Jin:2006if, Faulkner:2006ub} to approving 
\cite{Olmo:2006eh, Faraoni:2006hx} depending on the point of view of the author.
 
The essence of the discussion has been the question of validity of the Schwarzschild-de Sitter 
(SdS) metric as the correct metric in the solar system. The SdS metric is a vacuum 
solution to a large class of $f(R)$ theories of gravity but due to the higher-derivative 
nature of metric $f(R)$ theories, it is not unique. Other solutions can also be 
constructed in empty space, in the presence of matter and in a cosmological setting (see 
{\it eg.} \cite{cognola,Multamaki2,Multamaki}).

In light of recent literature \cite{Erickcek:2006vf,Chiba2,Jin:2006if}, the validity of
the solar system constraints has become clear and it is now understood that the
equivalent scalar-tensor theory results are valid in a particular limit that 
corresponds to the limit of light effective scalar in the
In terms of the $f(R)$ theory, this is equivalent to requiring that one can approximate
the trace of the field equations by Laplace's equation \cite{Chiba2}.
As a result, the often considered $R-\mu^4/R$ theory \cite{turner} (the CDTT model) is not
consistent with the Solar System constraints in this limit, if the $1/R$ term is to 
drive late time cosmological acceleration.

In \cite{Erickcek:2006vf} the CDTT model was considered by linearizing around a static de Sitter spacetime
and solving the trace equation in terms of $R(r)$, resulting in a spacetime outside the star where $\gamma=1/2$. 
This result was then generalized for a general $f(R)$ theory in \cite{Chiba2} by studying the space-time outside 
a spherical mass distribution and expanding $f(R)$ in terms of a perturbation in $R$. Again solving the 
trace equation leads to an outside solution with $\gamma=1/2$ as long as the effective scalar mass is light.
A somewhat different approach was followed
in \cite{kimmo}, where the trace equation was first written in terms $F(r)\equiv df/dR$ 
in the perturbative expansion. Solving the trace equation then leads to $\gamma=1/2$ outside the star.

In this paper we follow the latter approach by viewing $F$ along with the metric as independent functions.
By expanding all quantities in $G$ and solving the resulting equations inside and outside the star for a 
general $f(R)$ theory, we find that generally, $\gamma_{PPN}=1/2+\cO(G)$ outside the star and
the scalar curvature is $\cO(G^2)$ everywhere.
We also idenfity the first order correction to $\gamma_{PPN}$ and show that it cannot have a significant effect.
Only if initial conditions inside the star are fine-tuned such that the scalar curvature follows the matter density 
like in GR \cite{Hu2007, kimmo} can these bounds be evaded.

\section{$f(R)$ gravity formalism}

The action for $f(R)$ gravity is ($c=1$) 
\be{action}
S = \int{d^4x\,\sqrt{-g}\Big(\frac 1{16 \pi G} f(R)+{\cal{L}}_{m}\Big)}.
\ee
The field equations resulting in the so-called metric approach are
reached by variating with respect to $g_{\mu\nu}$: 
\be{eequs}
F(R) R_{\mu\nu}-\frac 12 f(R) g_{\mu\nu}-\nabla_\mu\nabla_\nu F(R)+g_{\mu\nu}\Box F(R)=8 
\pi G T^m_{\mu\nu},
\ee
where $T_{\mu\nu}^m$ is the standard minimally coupled stress-energy tensor
and $F(R)\equiv df/dR$.

Contracting the field equations and assuming that we can describe the
stress-energy tensor with a perfect fluid, we get
\be{contra}
F(R)R-2 f(R)+3\Box F(R)=8\pi G(\rho-3p).
\ee

In this letter we consider spherically symmetric static fluid configurations and adopt a metric, 
which reads in spherically symmetric coordinates as
\be{sphersym}
ds^2=B(r)dt^2-A(r)dr^2-r^2d\theta^2-r^2\sin^2\theta d\varphi^2.
\ee
By taking suitable linear combinations of the field equations they can be 
written in the following form:
\begin{widetext}
\bea{fieldequs0}
 \frac{F\,A'}{r\,A} + 
  \frac{F\,B'}{r\,B} + \frac{A'\,F'}{2\,A} + 
  \frac{B'\,F'}{2\,B} - F''
  & = & 8\,G\,\pi \,A\,(\rho +p)\\
  - \frac{F}{r^2}  + \frac{A\,F}{r^2}  + \frac{F\,A'}{2\,r\,A} + 
  \frac{F\,B'}{2\,r\,B} - 
  \frac{F\,A'\,B'}{4\,A\,B} - 
  \frac{F\,{B'}^2}{4\,{B}^2} - \frac{F'}{r} + 
  \frac{B'\,F'}{2\,B} + \frac{F\,B''}{2\,B}
  & = & 8\,G\,\pi \,A\,(\rho +p)\\
 A\,(2f(R)-R\,F(R))+ 
  \frac{6\,F'}{r} - \frac{3\,A'\,F'}{2\,A} + 
  \frac{3\,B'\,F'}{2\,B} + 3\,F''
   & = & -8\,G\,\pi \,A\,(\rho - 3 p),
  \eea
\end{widetext}
where prime indicates a derivation with respect to $r$, $'\equiv d/dr$
and we have written $f$ and $F$ as functions of the radial coordinate $r$
expect in combination $2f(R)-R\,F(R)$, which we will expand in terms of curvature $R$. 

The corresponding equation of continuity is
\be{cont}
\frac{p'(r)}{\rho(r)+p(r)}=-\frac 12 \frac{B'(r)}{B(r)}.
\ee 
When pressure is negligible, it is easy to see that $B$ must be a constant. This is, 
however, not acceptable and therefore an adequate perturbation expansion is needed. 

\section{Perturbative expansion and its solutions}

We expand the metric as well as $F$ with $G$ as an expansion parameter:
\bea{metexp}
A(r) & = & 1+G\, A_1(r)+ \cO(G^2),\nonumber\\
B(r) & = & B_0+G\, B_1(r)+ \cO(G^2),\nonumber\\
F(r) & = & F_0+G\, F_1(r)+ \cO(G^2),\nonumber\\
p(r) & = & p_0+G\, p_1(r)+ \cO(G^2).\nonumber 
\eea
Note, that we consider the density profile $\rho(r)$ to be a fixed function and
also that $B_0$ and $F_0$ are constants.
From the expansion of $A$ and $B$, one can also read out an expansion for $R$:
\be{Rexp}
R = R_0 + G R_1 + \cO(G^2). 
\ee
From the equation of continuity we see that at $\cO(G^0)$ pressure is constant
and exactly zero, $p_0=0$, simply because it vanishes in empty space. Therefore pressure effects 
are always $\cO (G^2)$ and do not contribute to the $\cO(G^1)$ expansion.
 
The $2f-FR$ term in the third field equations is crucial in determining the behaviour of 
the solution. In general, for a $f(R)$ dark energy model, this term is negligible and can 
be omitted, at least in the first order approximation. This is demonstrated explicitly 
for the CDTT in model and discussed more generally in \cite{kimmo}, where it is argued that 
the non-linear term is completely negligible, barring fine tuning. 
This argument is easily understandable in a general model since in the vacuum 
$2f-FR\sim G\rho_{DE}\ll G\rho$ for any stellar matter configuration. Note that this will 
in general be true also outside a stellar configuration as the dark matter will 
completely dominate over the cosmological term. Hence, in the trace equation, the 
non-linear terms can be dropped, unless the initial conditions are fine-tuned. We will 
return to the fine-tuned condition, or the Palatini limit \cite{Hu2007, kimmo}, later.

More formally, the same conclusion can be confirmed by using an expansion in $G$ for the non 
linear-terms as well:
$2f(R)-F(R)R = 2f(R_0)-F(R_0)R_0 + (F(R_0)- F'(R_0)R_0)R_1+ \cO (G^2)$
Evidently, the expansion point $R_0$ has to be such, that it corresponds to the correct
background of the theory, \ie $2f(R_0)-F(R_0)R_0=0$. Then 
expanding up to first order in $G$, the field equations are
\bea{fieldequsG}
\frac{F_0\,A_1'}{r} + \frac{F_0\,B_1'}{B_0\,r} - F_1'' & = & 8\,\pi \,\rho,\nonumber\\
\frac{F_0\,A_1}{r^2}-\frac{F_0\,A_1'}{2\,r} + \frac{F_0\,B_1'}{2\,B_0\,r} - \frac{F_1'}{r} 
+   \frac{F_0\,B_1''}{2\,B_0} & = & 8\,\pi \,\rho,\\
I_1 R_1+\frac{6\,F_1'}{r} + 3\,F_1'' & = & -8\,\pi \,\rho\nonumber,
\eea
where $I_1=F(R_0)- F'(R_0)R_0$ is a constant and $F_0= F(R_0)$.

The set of equations (\ref{fieldequsG}) can be straightforwardly solved leading to $\cO(G)$
functions:
\bea{solG}
F(r) & = & F_0-\frac 23 G\int_0^r \frac{m(r)}{r^2}\, dr\\
A(r) & = & 1+\frac{4G}{3F_0}\frac{m(r)}{r}\\
B(r) & = & B_0(1+\frac{8 G}{3F_0}\int_0^r \frac{m(r)}{r^2})\, dr,
\eea
where
\be{mdef}
m(r)\equiv \int_0^r 4\pi r^2\rho\, dr.
\ee 
Inserting this solution back to expression of the curvature scalar, we find that 
$R_1 =0$, {\it i.e.}, $R$ is $\cO(G^2)$. It is crucial that in deriving the solution (\ref{solG}), 
we have assumed that $A,\ B,\ F$ are regular at the origin. 

The PPN-parameter is now straightforwardly calculable:
\be{PPNG}
\gamma_{PPN}=\frac 12(1-r \frac{m'}{m}) + \frac{2 G}{3 F_0}
 \frac{\left( 2\,\int_0^r \frac{m}{r^2}\,dr + m' \right) \,
     \left( m - r\,m' \right)}{m}.
\ee
It is easy to see that, at the boundary of the star $\gamma_{PPN}\rar 1/2+\cO(G)$.
This behaviour was also observed in numerical studies \cite{kimmo,henttunen}.
From the first order correction one can furthermore conclude that if one wishes corrections 
to be effective at zeroth order, $F_0$ needs to be of order $G$.
However, looking at the continuity equation, Eq. (\ref{cont}), we find that
\be{contG}
-r^2p'=\frac 43\frac{G}{F_0}\rho\,m(r)+\cO(G^2).
\ee
Comparing this with the Newtonian result, $-r^2 p'=4G\rho m(r)$, we see that if 
$F_0\sim \cO(G)$, the effective Newton's constant is
orders of magnitude larger than the one in Newton's theory (or GR), resulting in stars with 
a completely different mass to radius relationship than
the one observed. Furthermore, from the continuity equation, we can read that unless 
$F_0\approx 4/3$ to a high precision, a star with the same
density profile, and hence total mass, will have a different radius than in GR. This 
behaviour was already observed in \cite{henttunen}.

In general the results described in this section will apply even when $2f-FR\sim R$, \ie 
when $f(R)=R+c_1 R^2$. Because in the approximation described
above, $R\sim \cO(G^2)$, it is easy to see that that this term will play no role in 
the trace equation. Similarly for higher order terms
in $R$. One can avoid the constraint only if $F$ has no $G$ order correction. In this case, 
the $\Box F$, term is negligible in trace equation and we recover the GR results, or the 
Palatini limit \cite{Hu2007, kimmo}. Alternatively, if one relaxes the regularity constraint of 
the metric at the origin,
one can also avoid the constraint as demonstrated in \cite{henttunen} for the CDTT model.

\subsection{Recovering the general relativity}

In the Palatini limit, where the trace-equation is similar than in the Palatini formalism,
the theory is fine-tuned so that $2f-FR\approx R\approx - 8\pi G\rho$ 
throughout (see \cite{kimmo} for a numerical example). This is the mechanism that allows one 
to construct
solutions that are consistent with solar system observations \cite{Hu2007}.
In the Palatini limit, the field equations read as
\bea{solP}
F(r) & \simeq & 1\\
A(r) & \simeq & 1+\frac{2Gm(r)}{r}\\
B(r) & \simeq & B_0(1+2 G \int_0^r \frac{m(r)}{r^2})\, dr.
\eea
The $\gamma_{PPN}$ parameter is easily calculable: 
\be{PPNP}
\gamma_{PPN} \simeq 1-r \frac{m'}{m}+\cO(G).
\ee
Therefore, in this limit $\gamma_{PPN}\rar 1$ at the surface of the star. However, as shown in 
\cite{kimmo} for the CDTT model, this limit can be unstable time leading to the Dolgov-Kawasaki 
instability \cite{dolgov}.  Stable theories are considered in \cite{Hu2007} and 
studied analytically in \cite{new1}.

\section{Discussion and Conclusions}

In this letter we have considered a general metric $f(R)$ theory in the presence of matter by
analyzing the field equations by perturbative means in linear order in the Newton's constant $G$.
We have shown explicitly that for a typical star, any modification of gravity from GR will 
naturally lead to physically unacceptable value $\gamma_{PPN}=1/2$. This places a very strong
constraint on any $f(R)$ theory, in particular when acting as a dark energy candidate. 
Furthermore, even if the gravity theory 
is not motivated by cosmology, but by other arguments, such as quantum gravity, the 
presence of non-linear terms can still lead to a space-time inconsistent with observations.

In this order of perturbation theory we can recover the observationally acceptable space-time,
only when $F=df/dR$ has no order $G$ correction. Such a constraint indicates
fine-tuning in the initial values of the solution so that one remains in the high curvature limit,
$R\sim G\rho$ throughout. However, the stability of such a fine-tuned solution may be problematic
\cite{kimmo}, although possible to obtain \cite{Hu2007,new1,new2}.

Since our analysis is of order $\cO(G^1)$, further study on the system, in 
particular second order perturbations in $G$, may affect the conclusions. Indeed, our 
analysis shows that the first order perturbation theory is essentially independent on 
the details of the underlying $f(R)$ theory. The only piece of information used was the 
knowledge that there are higher order derivatives in the equations of motion, \ie 
that the theory is not GR. New effects may appear in higher order perturbation theory, where 
finally the dependence on the functional form of $f(R)$ should become evident. However, our results suggest that
unless the solution is fine-tuned so that $R\sim G\rho$ throughout the mass distribution, 
a naive modification where a small correction is added to the Einstein-Hilbert action to solve the dark energy problem 
is not likely to pass the solar system constraints.

\acknowledgments
TM is supported by the Academy of Finland. 




\begin{thebibliography}{X}








\bibitem{turner}  S.~M.~Carroll, V.~Duvvuri, M.~Trodden and M.~S.~Turner,  
Phys.\ Rev.\ D {\bf 70}, 043528 (2004).  

\bibitem{turner2}  S.~M.~Carroll, A.~De Felice, V.~Duvvuri, D.~A.~Easson, M.~Trodden 
and M.~S.~Turner, Phys.\ Rev.\ D {\bf 71}, 063513 (2005).  

\bibitem{allemandi}  G.~Allemandi, A.~Borowiec and M.~Francaviglia,  
Phys.\ Rev.\ D {\bf 70}, 103503 (2004).  

\bibitem{meng}  X.~Meng and P.~Wang,  
Class.\ Quant.\ Grav.\  {\bf 21}, 951 (2004). 

\bibitem{nojiri3}  S.~Nojiri and S.~D.~Odintsov,  
Phys.\ Rev.\ D {\bf 68}, 123512 (2003).

\bibitem{cappo1} S.~Capozziello, Int.\ J.\ Mod.\ Phys.\ D {\bf 11}, 
483 (2002).  

\bibitem{nojiri2}  S.~Nojiri and S.~D.~Odintsov,  
Phys.\ Lett.\ B {\bf 576}, 5 (2003).  


\bibitem{woodard}  R.~P.~Woodard, 
arXiv:astro-ph/0601672.

\bibitem{odintsov} For dicussion of other generalized theories, see {\it e.g.}:
 S.~Nojiri and S.~D.~Odintsov, arXiv:hep-th/0601213.

\bibitem{cognola}  G.~Cognola, E.~Elizalde, S.~Nojiri, S.~D.~Odintsov and S.~Zerbini,  
JCAP {\bf 0502}, 010 (2005) [arXiv:hep-th/0501096].  

\bibitem{dolgov}  A.~D.~Dolgov and M.~Kawasaki,  
Phys.\ Lett.\ B {\bf 573}, 1 (2003).  

\bibitem{soussa}  M.~E.~Soussa and R.~P.~Woodard, 
 Gen.\ Rel.\ Grav.\  {\bf 36}, 855 (2004)  [arXiv:astro-ph/0308114].

\bibitem{faraoni}  V.~Faraoni,  
Phys.\ Rev.\ D {\bf 72}, 124005 (2005)  [arXiv:gr-qc/0511094].  

\bibitem{chiba}  T.~Chiba,  
Phys.\ Lett.\ B {\bf 575}, 1 (2003).

\bibitem{confprobs}  E.~E.~Flanagan, Class.\ Quant.\ Grav.\  {\bf 21}, 417 (2003);
Class.\ Quant.\ Grav.\  {\bf 21}, 3817 (2004);
G.~Magnano and L.~M.~Sokolowski,
Phys.\ Rev.\ D {\bf 50}, 5039 (1994).
\bibitem{Clifton}  T.~Clifton and J.~D.~Barrow,
 Phys.\ Rev.\ D {\bf 72}, 103005 (2005)  [arXiv:gr-qc/0509059].

\bibitem{Bean:2006up}  R.~Bean, D.~Bernat, L.~Pogosian, A.~Silvestri and M.~Trodden,  
arXiv:astro-ph/0611321.  

\bibitem{Song:2006ej}  Y.~S.~Song, W.~Hu and I.~Sawicki,  
arXiv:astro-ph/0610532.  

 \bibitem{Song2}
  Y.~S.~Song, H.~Peiris and W.~Hu,
  arXiv:0706.2399 [astro-ph].
  
  \bibitem{Sawicki2007}
  I.~Sawicki and W.~Hu,
  Phys.\ Rev.\  D {\bf 75}, 127502 (2007)
  [arXiv:astro-ph/0702278].
  
  \bibitem{Hu2007}
  W.~Hu and I.~Sawicki,
  arXiv:0705.1158 [astro-ph].
  
  
\bibitem{Appleby}
  S.~A.~Appleby and R.~A.~Battye,
  arXiv:0705.3199 [astro-ph].

\bibitem{damour}
  T.~Damour and G.~Esposito-Farese, 
 Class.\ Quant.\ Grav.\  {\bf 9}, 2093 (1992). 

\bibitem{magnano}
G.~Magnano and L.~M.~Sokolowski,
Phys.\ Rev.\ D {\bf 50}, 5039 (1994)  [arXiv:gr-qc/9312008].

\bibitem{olmo}  G.~J.~Olmo, 
 Phys.\ Rev.\ Lett.\  {\bf 95}, 261102 (2005)  [arXiv:gr-qc/0505101].  


\bibitem{ppnok}  S.~Capozziello, A.~Stabile and A.~Troisi,
  arXiv:gr-qc/0603071; 
 G.~Allemandi, M.~Francaviglia, M.~L.~Ruggiero and A.~Tartaglia,
  Gen.\ Rel.\ Grav.\  {\bf 37}, 1891 (2005)  [arXiv:gr-qc/0506123];
 T.~P.~Sotiriou,
  arXiv:gr-qc/0507027.  



\bibitem{Erickcek:2006vf}  A.~L.~Erickcek, T.~L.~Smith and M.~Kamionkowski,
 arXiv:astro-ph/0610483.  

\bibitem{Chiba2}  T.~Chiba, T.~L.~Smith and A.~L.~Erickcek,  
arXiv:astro-ph/0611867.  

\bibitem{Jin:2006if}  X.~H.~Jin, D.~J.~Liu and X.~Z.~Li,  
arXiv:astro-ph/0610854.  

\bibitem{Faulkner:2006ub}
  T.~Faulkner, M.~Tegmark, E.~F.~Bunn and Y.~Mao,  
arXiv:astro-ph/0612569.  

\bibitem{Faraoni:2006hx}
  V.~Faraoni,
  Phys.\ Rev.\  D {\bf 74}, 023529 (2006)
  [arXiv:gr-qc/0607016].

\bibitem{Olmo:2006eh}
  G.~J.~Olmo,
  Phys.\ Rev.\  D {\bf 75}, 023511 (2007)
  [arXiv:gr-qc/0612047].

\bibitem{Multamaki}  T.~Multamaki and I.~Vilja,
Phys.\ Rev.\ D {\bf 73}, 024018 (2006)  [arXiv:astro-ph/0506692].

\bibitem{Multamaki2}  T.~Multamaki and I.~Vilja,  
Phys.\ Rev.\ D {\bf 74}, 064022 (2006)  [arXiv:astro-ph/0606373].  

\bibitem{kimmo}
  K.~Kainulainen, J.~Piilonen, V.~Reijonen and D.~Sunhede,
  Phys.\ Rev.\  D {\bf 76}, 024020 (2007)
  [arXiv:0704.2729 [gr-qc]].

\bibitem{henttunen}
  K.~Henttunen, T.~Multamaki and I.~Vilja,
  arXiv:0705.2683 [astro-ph].
  
  
  
\bibitem{new1}
S. Nojiri, S. D. Odintsov, Phys.Lett.B652, 343-348 (2007) [arXiv:0706.1378 [hep-th]].
 
\bibitem{new2}
  S.~Tsujikawa,
  arXiv:0709.1391 [astro-ph].

\bibitem{Nojiri:2007as}
  S.~Nojiri and S.~D.~Odintsov,
  arXiv:0707.1941 [hep-th].

\bibitem{Starobinsky:2007hu}
  A.~A.~Starobinsky,
  arXiv:0706.2041 [astro-ph].

 

\end{thebibliography}
\end{document}